\documentclass[12pt]{article}
\usepackage{a4}
\usepackage{amsfonts,amssymb,amsmath}
\usepackage{graphicx}
\usepackage{cite}

\begin{document}

\title{ Gauge invariance and classical dynamics of noncommtative particle
theory}
\author{D.M. Gitman\thanks{e-mail: gitman@dfn.if.usp.br}, V.G.
Kupriyanov\thanks{e-mail: vladislav.kupriyanov@gmail.com} \\ Instituto de F\'{\i}sica, Universidade de S\~{a}o Paulo, Brazil}

\date{\today                                        }
\maketitle

\begin{abstract}
We consider a model of classical noncommutative particle in an external
electromagnetic field. For this model, we prove the existence of generalized
gauge transformations. Classical dynamics in Hamiltonian and Lagrangian form
is discussed, in particular, the motion in the constant magnetic field is
studied in detail.

\end{abstract}

\section{Introduction}

In the last decade, there has been a certain interest in considering
quantum-mechanical and field-theoretical models with noncommutative space-time
coordinates, see e.g. \cite{NCreviews} and \cite{NCQM} for reviews on
noncommutativity in QFT and QM, respectively. The noncommutative space can be
realized by the coordinates $\hat{x}^{i},$ satisfying commutation relations
$\left[  \hat{x}^{i},\hat{x}^{j}\right]  =i\theta^{ij},$ where $\theta^{ij\,}$
is an antisymmetric constant matrix. Classical actions of field theories on a
noncommutative space-time can be written as some modified classical actions
already on the commutative space-time, using the Weyl-Moyal correspondence
\cite{Moyal}. Similar possibility exists in the case of a finite-dimensional
theory (mechanics). The noncommutativity of position coordinates can be
obtained as a consequence of a canonical quantization of dynamical models
\cite{Lukierski1}-\cite{DN}. For example, the canonical quantization of the
classical theory with first-order action
\begin{align}
S_{NC}  &  =S_{H}+S_{\theta},\label{1}\\
S_{H}  &  =\int dt\left[  p_{j}\dot{x}^{j}-H\left(  p,x\right)  \right]
,\ \ S_{\theta}=\int dtp_{i}\theta^{ij}\dot{p}_{j}/2,\nonumber
\end{align}
leads to a quantum theory (which is called noncommutative quantum mechanics
(NCQM)) with the commutation relations%
\begin{equation}
\left[  \hat{x}^{i},\hat{x}^{j}\right]  =i\theta^{ij}\,,\;\left[  \hat{x}%
^{i},\hat{p}_{j}\right]  =i\delta_{j}^{i}\,,\;\left[  \hat{p}_{i},\hat{p}%
_{j}\right]  =0\,, \label{2}%
\end{equation}
and with the quantum Hamiltonian $H\left(  \hat{x},\hat{p}\right)  $. The
model of noncommutative particle (\ref{1}) was proposed in \cite{Duval}, see
also \cite{Deriglazov}. In fact, $S_{H}$ is the ordinary Hamiltonian action
and $S_{\theta}$ is responsible for noncommutativity.

As it was already mentioned, NCQM has been studied extensively \cite{NCQM},
and many calculations on the base of the theory were performed to find the
upper bound on the noncommutativity parameter $\theta.$ However, there remain
some open questions in classical mechanics of noncommutative particle, for
example, the problem of gauge invarience with respect to the external
electromagnetic field. It is known that NCQM with external electromagnetic
field is invariant under the noncommutative $U\left(  1\right)  $ gauge group,
which is $U_{\star}\left(  1\right)  ,$ see \cite{Chaichian}. This fact may
serve as an indication that there exist a classical version of such
transformations. In fact, the problem is closely related to the problem of
introducing the interaction with the Abelian gauge field in the classical
models of noncommutative particle (\ref{1}), see e.g. \cite{Lukierski}.

As it is now known, there exist two ways of introducing potentials $A_{\mu
}(x)=\left(  A_{i}(x),A_{0}(x)=\varphi\left(  x\right)  ,\ i=1,...,n\right)  $
of the external electromagnetic field in the theory, which correspond to two
different actions $S_{NC}^{1}=S_{H}^{1}+S_{\theta}$ of the Duval-Horvathy
model \cite{Duval}, and $S_{NC}^{2}=S_{H}^{2}+S_{\theta}$ of the Deriglazov
model \cite{Deriglazov}, where%
\begin{align}
S_{H}^{1}  &  =\int dt\left[  \left(  p_{j}+eA_{j}\left(  x\right)  \right)
\dot{x}^{j}-\frac{1}{2}p^{2}-e\varphi(x)\right]  ,\label{3}\\
S_{H}^{2}  &  =\int dt\left[  p_{j}\dot{x}^{j}-\frac{1}{2}\left[  p_{i}%
-eA_{i}(x)\right]  ^{2}-e\varphi(x)\right]  \label{4}%
\end{align}

The action $S_{NC}^{1}$, by the construction, is invariant under the $U\left(
1\right)  $ gauge transformations: $\delta A_{i}=\partial_{i}f\left(
x\right)  $ and the particle momenta have not to be changed. Classical
equations of motion describing dynamics of noncommutative particle in
Duval-Horvathy model were investigated in details in \cite{Plyuschay}.
Hamiltonization of the theory with the action $S_{NC}^{1}$ leads to the
following Dirac brackets between the phase space variables $x^{i},p_{j}$:%
\begin{align}
\left\{  x^{i},x^{j}\right\}  _{D}  &  =\varepsilon^{ij}\theta d,\ \ \left\{
x^{i},p_{j}\right\}  _{D(\Phi)}=\delta_{j}^{i}d,\ \ \nonumber\\
\left\{  p_{i},p_{j}\right\}  _{D}  &  =\varepsilon_{ij}eBd,\ \ d=1-e\theta
B\left(  x\right)  , \label{6}%
\end{align}
where $B\left(  x\right)  =\partial_{1}A_{2}-\partial_{2}A_{1}$ is the
magnetic field. After quantization they determine the commutation relations
between coordinates and momenta:%
\begin{equation}
\left[  \hat{x}^{i},\hat{x}^{j}\right]  =i\theta\varepsilon^{ij}d,\ \ \left[
\hat{x}^{i},\hat{p}_{j}\right]  =i\delta_{j}^{i}d,\ \ \left[  \hat{p}_{i}%
,\hat{p}_{j}\right]  =ieB\varepsilon_{ij}d. \label{7}%
\end{equation}
As it was first mentioned in \cite{GK09} the space noncommutativity depends on
magnetic field $B\left(  x\right)  $ and this is not a case of usually
considered noncommutativity with constant $\theta.$ The canonical quantization
of the Deriglazov model leads to the space noncommutativity with a constant
$\theta.$ That is why we concentrate our attention on the Deriglazov model in
what follows.

In spite of the fact that the action (\ref{4}) is invariant under the standard
gauge transformations:%
\begin{equation}
\delta A_{i}=\partial_{i}f\left(  x\right)  ,\delta p_{i}=e\partial
_{i}f\left(  x\right)  , \label{5}%
\end{equation}
the complete action $S_{NC}^{2}$ is not, due to the term $S_{\theta}$. In the
work \cite{Lukierski}, on the example of planar particle, $n=2$, only
infinitesimal local transformations%
\begin{align}
\delta x^{i}  &  =-e\theta\varepsilon^{ij}\partial_{j}\Lambda\left(  x\right)
,\ \ \delta p_{i}=e\partial_{i}\Lambda\left(  x\right)  ,\nonumber\\
\delta A_{i}  &  =A_{i}^{\prime}\left(  x+\delta x\right)  -A_{i}\left(
x\right)  =\partial_{i}\Lambda\left(  x\right)  , \label{5a}%
\end{align}
were constructed, which preserve simplectic structure of $S_{NC}^{2}$, and
change corresponding Lagrangian on the total time derivative.

In the present article we demonstrate the existance of generalized gauge
transformations for the Deriglazov model. These transformations are
deformation in $\theta$ of the gauge transformations (\ref{5}). In the first
order in $\theta$ they coincide with (\ref{5a}). After quantization the
generalized gauge transformations lead to the gauge group of NCQM, see
\cite{Chaichian}. Then, we consider classical dynamics of the model in the
configuration space, and a possibility to construct a Lagrangian second-order
action which is equivalent to the Hamiltonian first-order action (\ref{1}).
The general consideration is illustrated by an example of the noncommutative
charged particle in a constant magnetic field.

\section{Generalized gauge transformations}

The action $S_{NC}^{2}$ can be written as follows:%
\begin{align}
S_{NC}^{2}  &  =\int dtL_{H}^{\theta}~,\ \ \ L_{H}^{\theta}=L_{1}%
-H,\ \ \nonumber\\
L_{1}  &  =p_{j}\dot{x}^{j}+\frac{1}{2}p_{i}\theta^{ij}\dot{p}_{j}%
~,\ H=\frac{1}{2}\left(  p_{i}-eA_{i}(x)\right)  ^{2}+e\varphi\left(
x\right)  . \label{8}%
\end{align}
The simplectic structure (Poisson brackets) corresponding to this first-order
action is:%

\begin{equation}
\left\{  x^{i},x^{j}\right\}  =\theta^{ij},\ \ \left\{  x^{i},p_{j}\right\}
=\delta_{j}^{i},\ \ \left\{  p_{i},p_{j}\right\}  =0. \label{9}%
\end{equation}

Below, we are going to construct an explicit form of the generalized gauge
transformations. To this end, first, we introduce the following
transformations:%
\begin{equation}
\delta x^{i}=K^{i}\left(  x\right)  ,\ \ \delta p_{i}=J_{i}\left(  x\right)
,\ \ J_{i}\left(  x\right)  =e\partial_{i}f\left(  x\right)  +O\left(
\theta\right)  , \label{11}%
\end{equation}
which should leave the simplectic structure (\ref{9}) invariant. That is, new
coordinates and momenta%
\begin{equation}
x^{\prime i}=x^{i}+K^{i}\left(  x\right)  ,\ \ p_{i}^{\prime}=p_{i}%
+J_{i}\left(  x\right)  \label{12}%
\end{equation}
must have the same Poisson brackets. From this condition one obtains equations
on the functions $K^{i}\left(  x\right)  $ and $J^{i}\left(  x\right)  $:%
\begin{align}
&  \theta^{il}\partial_{l}K^{j}-\theta^{jl}\partial_{l}K^{i}+\left\{
K^{i},K^{j}\right\}  =0,\nonumber\\
&  \theta^{il}\partial_{l}J_{j}-\partial_{j}K^{i}+\left\{  K^{i}%
,J_{j}\right\}  =0,\ \partial_{i}J_{j}-\partial_{j}J_{i}+\left\{  J_{i}%
,J_{j}\right\}  =0, \label{13a}%
\end{align}
where the Poisson brackets between two functions of coordinates are determined
as%
\[
\left\{  F,G\right\}  =\left(  \partial_{k}F\right)  \theta^{kl}\left(
\partial_{l}G\right)  .
\]
If $K^{i}=-\theta^{il}J_{l}$ then two first equations (\ref{13a}) are just the
consequences of the third one.

Thus, to find (\ref{11}) we have to solve the following differential equation:%
\begin{equation}
\partial_{j}J_{i}-\partial_{i}J_{j}=\left\{  J_{i},J_{j}\right\}  , \label{15}%
\end{equation}
with the condition that $J_{i}\left(  x\right)  =e\partial_{i}f\left(
x\right)  +O\left(  \theta\right)  $. The solution of this equation can be
found as a perturbative series in $\theta$, and has the form%
\begin{equation}
J_{i}\left(  x\right)  =\sum_{m=0}^{\infty}\frac{e^{m+1}}{(m+1)!}\underset
{m}{\underbrace{\{...\{}}\partial_{i}f,\underset{m}{\underbrace{f\},...,f\}}%
}=\sum_{m=0}^{\infty}J_{i}^{m}\left(  x\right)  , \label{16}%
\end{equation}
where\footnote{By the construction, the function $J_{i}^{m}\left(  x\right)  $ is of the
$m$-th order in $\theta$.} %
\begin{align}
J_{i}^{m}\left(  x\right)   &  =\frac{e}{m+1}\left\{  J_{i}^{m-1},f\right\}
,\ \ \ m\geq1,\label{17}\\
J_{i}^{0}\left(  x\right)   &  =e\partial_{i}f\left(  x\right)  .\nonumber
\end{align}

Let us prove it by the induction. One can easily
verify that%
\[
J_{i}^{1}\left(  x\right)  =\frac{e^{2}}{2}\left\{  \partial_{i}f,f\right\}
=\frac{e}{2}\left\{  J_{i}^{0},f\right\}
\]
is the solution of the equation (\ref{15}) in the first order in $\theta$. We
should prove that if $J_{i}^{m}\left(  x\right)  $\ is the solution of this
equation in the $m$-th order, i.e.,%
\begin{equation}
\partial_{j}J_{i}^{m}-\partial_{i}J_{j}^{m}=\sum_{l=0}^{m-1}\left\{
J_{i}^{m-1-l},J_{j}^{l}\right\}  , \label{17a}%
\end{equation}
holds, then the solution in the order $m+1$ is:%
\begin{equation}
J_{i}^{m+1}\left(  x\right)  =\frac{e}{m+2}\left\{  J_{i}^{m},f\right\}  .
\label{18}%
\end{equation}
Let us consider the following quantity%
\[
I_{ij}=\left(  m+2\right)  \left(  \partial_{j}J_{i}^{m+1}-\partial_{i}%
J_{j}^{m+1}\right)  =e\left(  \partial_{j}\left\{  J_{i}^{m},f\right\}
-\partial_{i}\left\{  J_{j}^{m},f\right\}  \right)  .
\]
With the help of (\ref{17a}), it can be rewritten as%
\[
I_{ij}=\sum_{l=0}^{m-1}e\left\{  \left\{  J_{i}^{m-1-l},J_{j}^{l}\right\}
,f\right\}  +\left\{  J_{i}^{m},e\partial_{j}f\right\}  +\left\{
e\partial_{i}f,J_{j}^{m}\right\}  .
\]
Using the Jacobi identity and (\ref{17}), we reduce $I_{ij}$ to the following
form%
\begin{align*}
&  \sum_{l=0}^{m-1}\left[  \left(  m-l+1\right)  \left\{  J_{i}^{m-l}%
,J_{j}^{l}\right\}  +\left(  l+2\right)  \left\{  J_{i}^{m-1-l},J_{j}%
^{l+1}\right\}  \right]  +\left\{  J_{i}^{m},J_{j}^{0}\right\}  +\left\{
J_{i}^{0},J_{j}^{m}\right\} \\
&  =\sum_{l=0}^{m-1}\left[  \left(  m-l+1\right)  \left\{  J_{i}^{m-l}%
,J_{j}^{l}\right\}  +\left(  l+1\right)  \left\{  J_{i}^{m-l},J_{j}%
^{l}\right\}  \right]  +\left(  m+2\right)  \left\{  J_{i}^{0},J_{j}%
^{m}\right\} \\ & =\left(  m+2\right)  \sum_{l=0}^{m}\left\{  J_{i}^{m-l}%
,J_{j}^{l}\right\}  ,
\end{align*}
and prove therefore that%
\[
\partial_{j}J_{i}^{m+1}-\partial_{i}J_{j}^{m+1}=\sum_{m=0}^{m}\left\{
J_{i}^{m-1-m},J_{j}^{m}\right\}  .
\]
In turn, this means that (\ref{18}) is a solution of equation (\ref{15}) in
$\left(  m+1\right)  $-th order with respect to $\theta$.

Finally we obtain:%
\begin{align}
\delta x^{i}  &  =K^{i}\left(  x\right)  =-\sum_{m=1}^{\infty}\frac{e^{m}}%
{m!}\underset{m}{\underbrace{\{...\{}}x^{i},\underset{m}{\underbrace
{f\},...,f\}}},\nonumber\\
\delta p_{i}  &  =J_{i}\left(  x\right)  =\sum_{m=0}^{\infty}\frac{e^{m+1}%
}{(m+1)!}\underset{m}{\underbrace{\{...\{}}\partial_{i}f,\underset
{m}{\underbrace{f\},...,f\}}}. \label{19a}%
\end{align}

The invariance of the Hamiltonian $H$ from (\ref{8}) under the transformations
(\ref{19a})\ implies the generalized gauge transformation of the potential
$A_{\mu}(x)$:%
\begin{align}
A_{i} &  \rightarrow A_{i}^{\prime}(x^{i}+\delta x^{i})=A_{i}(x)+\frac{1}%
{e}\delta p_{i}~,\nonumber\\
\varphi &  \rightarrow\varphi^{\prime}\left(  x+\delta x^{i}\right)
=\varphi\left(  x\right)  .\label{20}%
\end{align}
An explicit form of the transformed potential $A_{i}^{\prime}(x)$ can be
obtained by iterating the relation%
\[
A_{i}^{\prime}\left(  x\right)  =A_{i}(x)+\frac{1}{e}\delta p_{i}-\sum
_{m=1}^{\infty}\frac{e^{m}}{m!}\partial_{i_{1}}...\partial_{i_{m}}%
A_{i}^{\prime}\left(  x\right)  \delta x^{i_{1}}...\delta x^{i_{m}}.
\]
Up to the first order, one can obtain
\[
A_{i}^{\prime}\left(  x\right)  =A_{i}(x)+\partial_{i}f+e\left\{
A_{i}+3/2\partial_{i}f,f\right\}  ,\ \ \varphi^{\prime}\left(  x\right)
=\varphi\left(  x\right)  +\left\{  \varphi,f\right\}  .
\]
That is,
\begin{align*}
\delta A_{i} &  =A_{i}^{\prime}\left(  x\right)  -A_{i}(x)=\partial
_{i}f+e\left\{  A_{i}+3/2\partial_{i}f,f\right\}  +o\left(  \theta\right)  ,\\
\delta\varphi &  =\varphi^{\prime}\left(  x\right)  -\varphi\left(  x\right)
=\left\{  \varphi,f\right\}  +o(\theta).
\end{align*}
Since both Poisson brackets (\ref{9}) and the Hamiltonian (\ref{8}) are
invariant under the transformations (\ref{19a}), (\ref{20}), the corresponding
classical dynamics%
\begin{equation}
\dot{x}^{i}=\left\{  x^{i},H\right\}  ,\ \ \dot{p}_{i}=\left\{  p_{i}%
,H\right\}  ,\label{29}%
\end{equation}
is invariant under these transformations.

Following \cite{Lukierski}, we introduce non-Abelian field strength%
\begin{equation}
F_{ij}^{\theta}=\left\{  p_{i}-eA_{i}(x),p_{j}-eA_{j}(x)\right\}
=\partial_{i}A_{j}-\partial_{j}A_{i}+e\left\{  A_{i},A_{j}\right\}  .
\label{21}%
\end{equation}
By the definition, it is invariant under the generalized gauge transformations
(\ref{19a}) and (\ref{20}). After the quantization the corresponding field
strength is determined by%
\begin{equation}
F_{ij}^{\star}=\partial_{i}A_{j}-\partial_{j}A_{i}+e\left[  A_{i}%
,A_{j}\right]  _{\star}\ , \label{22}%
\end{equation}
where $F_{ij}^{\star}$ is the strength tensor of a gauge field related to
non-Abelian $U_{\star}\left(  1\right)  $ group, the latter is the gauge group
of NCQM, see \cite{Chaichian}.

As an example, we consider the case $n=2,$ $\varphi=0,$ and $A_{i}=\left(
-By/2,Bx/2\right)  ,$ which corresponds to a planar particle in a constant
magnetic field. Let $f=Bxy/2,$ then (\ref{16}) reads,%
\begin{align*}
&  J_{1}=ay,\ \ J_{2}=bx,\ \\
&  a=\frac{eB}{2}-\frac{2-\sqrt{e^{2}B^{2}\theta^{2}+4}}{2\theta
},\ \ \ b=\frac{eB}{2}+\frac{2-\sqrt{e^{2}B^{2}\theta^{2}+4}}{2\theta}.
\end{align*}
Using formulas (\ref{19a}) and (\ref{20}), we find the following gauge
transformations%
\begin{align}
&  x\rightarrow x^{\prime}=\left(  1-\theta b\right)  x,\ \ y\rightarrow
y^{\prime}=\left(  1+\theta a\right)  y,\nonumber\\
&  p_{1}\rightarrow p_{1}^{\prime}=p_{1}+ay,\ \ \ p_{2}\rightarrow
p_{2}^{\prime}=p_{2}+bx,\nonumber\\
&  A_{1}\rightarrow A_{1}^{\prime}=\frac{2a-eB}{2e\left(  1+\theta a\right)
}y^{\prime},\ \ A_{2}\rightarrow A_{2}^{\prime}=\frac{2b+eB}{2e\left(
1-\theta b\right)  }x^{\prime}. \label{b2}%
\end{align}
One can easily verify that the corresponding variation of the Lagrangian
(\ref{8}) is reduced to a total derivative,
\[
\delta L_{H}^{\theta}=\frac{d}{dt}\left[  \frac{1}{2}\left(  a-b\right)
xy+\theta ap_{2}y-\theta bp_{1}x\right]  .
\]
In the limit $\theta\rightarrow0,$ transformations (\ref{b2}) are reduced to
the gradient\ gauge transformations $A_{i}\rightarrow A_{i}^{\prime}=\left(
0,Bx\right)  $.

\section{Dynamics in configuration space and Lagrangian action}

Considering the Deriglazov model, we introduce new variables: $\left(
x^{i},p_{i}\right)  $ $\rightarrow\left(  x^{i},\pi_{i}\right)  $, where
$\pi_{i}=p_{i}-eA_{i}$. Poisson brackets involving new variables are%

\begin{equation}
\left\{  x^{i},\pi_{j}\right\}  =\delta_{j}^{i}~-e\theta^{ik}\partial_{k}%
A_{j},\ \ \left\{  \pi_{i},\pi_{j}\right\}  =eF_{ij}^{\theta}~,\ \
\end{equation}
such that the equations of motion take the form%
\begin{align}
\dot{x}^{i}  &  =\left\{  x^{i},H\right\}  ~=\left(  \delta_{j}^{i}%
~-e\theta^{ik}\partial_{k}A_{j}\right)  \pi_{j}+e\theta^{ij}\partial
_{j}\varphi,\nonumber\\
\dot{\pi}_{i}  &  =\left\{  \pi_{i},H\right\}  ~=eF_{ij}^{\theta}\pi
_{j}-e\left(  \delta_{j}^{i}~-e\theta^{ik}\partial_{k}A_{j}\right)
\partial_{j}\varphi~,\nonumber\\
H  &  =\pi^{2}/2+\varphi\left(  x\right)  . \label{32}%
\end{align}

Excluding momenta $\pi_{i}$ from equations (\ref{32}), we obtain second-order
equations of motion for the coordinates $x^{i}$. For simplicity, let us set
$\varphi\left(  x\right)  =0$, and $A_{i}\left(  x\right)  $ to be an
arbitrary function of the coordinates. Then we obtain $\theta$-modified
Lorentz equations in the case under consideration,%
\begin{align}
&  \ddot{x}^{i}=F_{\theta}^{i}+\tilde{F}^{i}~,\nonumber\\
&  F_{\theta}^{i}=eF_{ij}^{\theta}\dot{x}^{j}~,\ \ \tilde{F}^{i}=e\theta
^{ki}\partial_{k}\partial_{l}A_{j}\left(  \delta_{m}^{j}~-e\theta^{jn}%
\partial_{n}A_{m}\right)  ^{-1}\dot{x}^{l}\dot{x}^{m}~. \label{33a}%
\end{align}

If $\theta=0,$ the equations are reduced to the ordinary Lorentz equations. If
$\theta\neq0$, the Lorentz force $F_{\theta}^{i}$ is changed, according to
(\ref{21}), and a new force $\tilde{F}^{i}$ proportional to square of
velocities appears .

In the case of linear potential $A_{i}$, the term $\tilde{F}^{i}$ vanishes and
$F_{ij}^{\theta}$ is just a constant. If $n=2$ and $A_{i}=\left(
-By/2,Bx/2\right)  $ (the above considered magnetic field), equations
(\ref{33a}) take the form:%
\begin{equation}
\ddot{x}=e\tilde{B}\dot{y}~,\ \ \ddot{y}=-e\tilde{B}\dot{x}~,\ \ \tilde
{B}=B\left(  1+e\theta B/4\right)  . \label{35a}%
\end{equation}
Its solutions were analyzed in \cite{Baldiotti}.

Now we set $A_{i}=0$ and $\varphi\left(  x\right)  $ to be an arbitrary
function. In this case equations (\ref{32}) yield%
\begin{equation}
\ddot{x}^{i}-e\theta^{ij}\partial_{j}\partial_{k}\varphi\dot{x}^{k}%
+e\partial_{i}\varphi=0~. \label{35b}%
\end{equation}

Considering $n=2$ and $\varphi=\omega^{2}\left(  x^{2}+y^{2}\right)  /2$, we
obtain:
\[
\ddot{x}-e\theta\omega^{2}\dot{y}+e\omega^{2}x=0~,\ \ \ddot{y}+e\theta
\omega^{2}x+e\omega^{2}y=0~.
\]
The latter equations coincide with equations of motion of a charge in a
constant magnetic field $B_{\theta}=\theta\omega^{2}$ and linear electric
field $\mathbf{E}=\left(  \omega^{2}x,\omega^{2}y\right)  $, i.e., in this
case noncommutativity is equivalent to the presence of a magnetic field.

If $\varphi=y^{2}/2$, equations (\ref{35b}) read%
\begin{equation}
\ddot{x}-e\theta\dot{y}=0~,\ \ \ddot{y}+ey=0~.\nonumber
\end{equation}
This is a second-order non-Lagrangian set of equations, which does not admit
an integrating multiplier, see \cite{GK2,GK}.

If $n=2,$ $A_{i}=\left(  -By/2,Bx/2\right)  $ and $\varphi=\omega^{2}\left(
x^{2}+y^{2}\right)  /2$, we have%
\[
\ddot{x}^{i}-e\left(  \tilde{B}+\theta\omega^{2}\right)  \varepsilon^{ij}%
\dot{x}^{j}+e\omega^{2}\left(  1+\frac{e^{2}\theta^{2}B^{2}}{2}\right)
x^{i}=0~.
\]
These equations coincide with equations of motion of a charge in a constant
magnetic field $B_{\theta}=\tilde{B}+\theta\omega^{2}$ and a linear electric
field $\mathbf{E}=\omega^{2}\left(  1+e^{2}\theta^{2}B^{2}/2\right)  \left(
x,y\right)  $. For $\theta=-4B/\left(  4\omega^{2}+eB^{2}\right)  $, the
effective magnetic field $B_{\theta}$ disappears.

In fact, the noncommutative particle action (\ref{1}) is a first-order action,
and can be treated as a Hamiltonian action. To construct a second-order
Lagrangian formulation, we pass to Darboux coordinates. Namely, we change the
variables as follows: $(x^{i},p_{i})\rightarrow(q^{i},p_{i})$, where
\begin{equation}
q^{i}=x^{i}+\frac{1}{2}\theta^{ij}p_{j}~. \label{36}%
\end{equation}
In the new variables, the action (\ref{1}) takes the form%
\begin{equation}
S^{\theta}\left[  q,p\right]  =\int dt\left[  p_{i}\dot{q}^{i}-H\left(
q^{i}-\theta^{ij}p_{j}/2,p_{i}\right)  \right]  ~, \label{37}%
\end{equation}
where $H\left(  x,p\right)  $ is defined in (\ref{8}). From the equations%
\begin{equation}
\frac{\delta S^{\theta}\left[  q,p\right]  }{\delta p_{i}}=0\Rightarrow\dot
{q}^{i}=\frac{\partial H}{\partial p_{i}}\ \label{38}%
\end{equation}
one can express the momenta $p_{i}$ via coordinates $q^{i}$ and velocities
$\dot{q}^{i}$:%
\begin{align}
&  p_{i}=\dot{q}^{i}+eA_{i}\left(  q\right)  -e\partial_{j}A_{i}\left(
q\right)  \theta^{jk}\left[  \dot{q}^{k}+eA_{k}\left(  q\right)  \right]
\label{39}\\
&  +e\theta^{ij}\partial_{j}\varphi\left(  q\right)  +e\theta^{ij}\partial
_{j}A_{k}\left(  q\right)  \dot{q}^{k}+o\left(  \theta\right)  .\nonumber
\end{align}
Substituting (\ref{39}) into (\ref{37}), we obtain a second-order Lagrangian
action $S_{L}^{\theta}$ that does not contain any momenta,%
\begin{align*}
&  S_{L}^{\theta}=\int dtL^{\theta},\ \ L^{\theta}=\frac{1}{2}\dot{q}%
^{2}+eA_{i}\dot{q}^{i}-e\varphi\left(  q\right) \\
&  -e\dot{q}^{i}\partial_{j}A_{i}\theta^{jk}\left(  \dot{q}^{k}+eA_{k}\right)
-e^{2}\partial_{i}\varphi\theta^{ij}A_{j}+o\left(  \theta\right)  .
\end{align*}
Such a form of noncommutative particle actions can be useful both for
constructing Lagrangian path integrals in noncommutative quantum mechanics,
and for searching integrals of motion. E.g., having the Lagrangian $L^{\theta
}$, we easily obtain the conserved energy%
\[
E_{\theta}=\frac{1}{2}\dot{q}^{2}+e\varphi\left(  q\right)  +e^{2}\partial
_{i}\varphi\theta^{ij}A_{j}-e\dot{q}^{i}\frac{\partial A_{i}}{\partial q^{j}%
}\theta^{jk}\dot{q}^{k}+o\left(  \theta\right)  .
\]

Let us consider the above construction for a specific case where
$n=2,\ A_{i}=\left(  -By/2,Bx/2\right)  $, and $\varphi=\omega^{2}\left(
x^{2}+y^{2}\right)  /2$. In this case equations (\ref{38}) can be solved
exactly. Thus, we obtain%
\[
L^{\theta}=\varkappa\left[  \left(  \dot{q}_{x}^{2}+\dot{q}_{y}^{2}\right)
+e\left(  \tilde{B}+\omega^{2}\theta/4\right)  \left(  q_{x}\dot{q}_{y}%
-q_{y}\dot{q}_{x}\right)  -e\omega^{2}\left(  q_{x}^{2}+q_{y}^{2}\right)
\right]  ~,
\]
where%
\[
\varkappa=\left(  2+e^{2}B^{2}\theta^{2}/8+eB\theta+e\omega^{2}\theta
^{2}/2\right)  ^{-1}~.
\]
The corresponding $\theta$-modification of the usual conserved energy
$E_{0}=\left(  \dot{q}_{x}^{2}+\dot{q}_{y}^{2}\right)  /2+e\omega^{2}\left(
q_{x}^{2}+q_{y}^{2}\right)  /2$ is reduced to a multiplication by a factor,
$E_{\theta}=2\varkappa E_{0}.$

\section*{Acknowledgements}

We are grateful to Dmitri Vassilevich for fruitful discussions and to Peter
Horvathy for correspondence. The authors thanks FAPESP
and CNPq for support.

\end{document}